\documentclass[11pt]{article}
\title{NJL model in Four Dimension at Finite Temperature, Chemical Potential
and Curvature}
\author{Ashok Goyal\thanks{E--mail : agoyal@ducos.ernet.in}, 
and
Meenu Dahiya\thanks{E--mail : meenu@ducos.ernet.in}  \\
	{\em Department of Physics and Astrophysics,} \\
	{\em University of Delhi, Delhi-110 007, India.} \\
        {\em InterUniversity Centre for Astronomy and Astrophysics,} \\
        {\em Ganeshkhind, Pune 411007 , India.} \\
        }

\begin{document}
\maketitle
\large
\begin{abstract}
Two flavor Nambu-Jona-Lasinio model with N components is studied 
in curved space time at finite temperature and density in the leading 
1/N expansion. In four space time dimension the model exhibits first 
order phase transition for positive curvature. Whereas in flat space 
an increase in temperature induces a second order phase transition even 
at finite density, in curved space the transition becomes first order.
We obtain the phase boundary in density-temperature plane and exhibit
its behavior with changing curvature. A three dimensional plot of the
phase boundary in $\mu$-T-R space is drawn.
\end{abstract}
\pagebreak
\begin{section} {Introduction}
\indent Spontaneous symmetry breaking (SSB) is one of the most important
concepts of all unified gauge theories i.e  the underlying gauge symmetry
is larger than that of the vacuum. Of particular interest is the 
expectation that at high temperatures, symmetries that are spontaneously broken
today are restored and that during the evolution, the universe passed through
a series of phase transition evolving from a higher symmetric phase to a
lower symmetric phase associated with the spontaneous breakdown of gauge or 
perhaps global symmetry. It is very instructive to investigate how the phase 
transition takes place in quantum field theory in the environment of the 
early universe where the temperature, density, primordial external electromagnetic 
field and external gravity may all play important role \cite{linde}.
  
 In the standard grand unified theories, the Higgs field plays a most 
fundamental role. It is an elementary scalar field with self coupling and has
Yukawa couplings with all other fundamental fields-fermions and gauge bosons
giving them masses through its non-vanishing vacuum expectation value. There also
exist theories such as technicolour model of electro-weak theory \cite{weinberg}
 where the 
Higgs field may be considered as a composite field of some more fundamental
fermion fields. QCD is an another example of quantum field theory which is 
invariant under chiral transformations at the lagrangian level in the absence of 
quark mass matrix. However, the dynamics of the QCD are expected to be such 
that chiral symmetry is dynamically broken with the vacuum state having 
a nonzero quark-antiquark condensate $ \left\langle 0\left\vert\bar
qq\right\vert 0 \right\rangle $  and the Goldstone theorem then requries the 
existence of approximately massless  pseudo-scalar mesons in the hadron spectrum.
To study chiral phase transition in QCD we need a non-perturbative treatment
and a particularly attractive framework to study such systems is the Nambu-Jona 
-Lasinio (NJL) model \cite{nambu}.
 Linear sigma model is another such model that is amenable
to such a study.
       The NJL model ( for a recent review see \cite{klevansky} and references 
therein )
provides a useful framework for studying dynamical chiral symmetry breaking
non-perturbatively in the 1/N expansion, where starting from massless quarks,
 a vacuum condensate  $ \left\langle \bar qq \right\rangle \neq$ appears and
a dynamical fermion mass is generated with the breakdown of chiral symmetry
in the ground state. Further, NJL model can also be considered  as a prototype
of composite Higgs model and we will use it in the present study to investigate
the breakdown of higher symmetry in the early universe.

The NJL model four fermion theory in flat space-time in arbitrary dimensions
has been studied \cite{inagaki1} using the 1/N expansion method. It is shown that chiral 
symmetry is restored in the theory under consideration for sufficiently high 
temperature or chemical potential. It is found that for space-time dimensions
$2 \leq D < 3$ both a first order and second order phase transition occur
 depending on the value of the four fermion coupling while 
for $3 \leq D < 4$  only the second order phase transition exists.
 Recently an investigation of the chiral phase structure of the four fermion theory in curved space-time 
has been made [6-13]. In particular the existence of a 
first order phase transition
induced by curvature from the chiral symmetric phase to the chiral non-symmetric 
one has been shown. Chiral symmetry structure with nontrivial topoligies 
\cite{muta1,elizalde3,kim}
under the combined action of magnetic fields and curved space-time has also been
studied \cite{geyer}.
  In the present paper we will investigate the D=4 NJL model in curved space-time
at finite temperature, density and curvature using the schwinger proper time 
method \cite{schwinger}. We will also take into account the effect of
 temperature and density
on the leading contribution on curvature. Linear Sigma model will be topic of our
subsequent study. The existence of first order phase transition at finite 
temperature and curvature is confirmed and the phase diagram in the curvature,
temperature and density plane is obtained.  
\end{section}
\begin{section}{Four Fermion Theory in Curved Space-time}
\indent The Nambu Jona Lasinio model in curved space time is defined by the action
\begin{equation}
S=\int d^D x \, \sqrt{(-g)} \,[\,\, i \,\bar{\psi} \gamma^\mu(x) \bigtriangledown_\mu \psi
+\frac{\lambda}{2 N} (\,\,{(\bar{\psi}\psi)^2+(\bar{\psi}i\, \gamma_5 
\,\vec{\tau} \,\psi)^2})\,\,]
\end{equation}
where $g$  is the determinant of the space time metric ,$\gamma^\mu(x)$ the
Dirac matrix in curved space-time, $\bigtriangledown_\mu \psi$ the covariant 
derivative of the fermion field $\psi$ and N is the number of colours, we take the number of flavours to be two.
We work in the scheme of the 1/N expansion and perform our calculations in the
leading order of the expansion.
For practical purposes it is more convenient to introduce the auxiliary
fields $\sigma$ and $\vec{\pi}$ and consider the equivalent action.
\begin{equation}
S=\int d^D x \sqrt{(-g)}\, [ i\, \bar{\psi} \gamma^\mu(x) 
\bigtriangledown_\mu \psi-\frac{ N}{2 \lambda} (\sigma^2+\pi^2)
- \bar{\psi}( \sigma+ i \,\gamma_5 \,\vec{\tau} \, \vec{\pi}) \psi ]
\end{equation}
Replacing $\sigma$ and $\vec{\pi}$  by the solutions of the Euler-Lagrangian 
equations arising from (2) we reproduce the action (1). This is because the fields 
$\sigma$ and $\vec{\pi}$ are not independent degrees of freedom in (2) and the
Euler-Langrangion equations for  $\sigma$ and $\vec{\pi}$ are infact constraint equations which fix $\sigma$ and $\vec{\pi}$ given $\psi$ and $\bar \psi$.

If a non vanishing vacuum expectation value is assigned to the auxiliary
field $\sigma$, then there appears a mass term for the fermion field
$\psi$ and the discrete chiral symmetry is eventually broken.

The effective potential (with N factored out) in the leading order 
of the 1/N expansion is then given by 
\begin{equation}
V(\sigma, \pi)=\frac{1}{2 \lambda} \,(\sigma^2+\vec\pi^2) + i \,Tr\, ln\left\langle 
x \left\vert\,\,(i \gamma^\mu(x) \bigtriangledown_\mu -( \sigma+ i\, \gamma_5 
\,\vec{\pi}) \,\,\right\vert x \right\rangle  
\end{equation}
 
In equation (3) the variables $\sigma$ and $\vec{\pi}$ are regarded as constant.
Using the schwinger proper time method \cite{schwinger} we rewrite the second 
term on the right  hand side of equation (4) as
\begin{equation}
V(\sigma, \pi)= \frac{1}{2 \lambda} (\sigma^{2}+\pi^{2}) + i \,Tr \,  ln \,
S\,(x,x;s) \,\Bigg\vert_{s=\sigma+\,i \gamma_5\, \vec{\pi}}
\end{equation}
 
where the Green function S  defined by
\begin{equation}
S(x,y;s)= \left\langle x \left\vert\,\, (i \gamma^\mu(x) \bigtriangledown_\mu -s)^{-1}
 \right\vert y \right\rangle  
\end{equation}

is the solution of the equation
\begin{equation}
(i \gamma^\mu(x) \bigtriangledown_\mu -s) S(x,y;s)= \frac{1}{\sqrt{(-g(x))}} \delta^D(x-y)
\end{equation}
Thus the effective potential is described by the two point Green's function 
S(x,x;s) of the massive free fermion in curved spacetime.
Using the Green function obtained in the approximation of keeping only 
linear terms
in the curvature, Inagaki, Muta and Odintsov \cite{inagaki3}, obtained the 
effective potential which in D-dimensions read as follows:
\begin{eqnarray}
V( \sigma, 0) &=& \frac{\sigma^{2}}{2 \lambda} -i Tr  \int_{0}^{\sigma} d s \int
 \frac{d^D k}{(2 \pi)^4} [ (\gamma^{a}  k_{a}+s) \frac{1}{k^2-s^2}
\nonumber \\
&&-\frac{R}{12} (\gamma^{a}  k_{a}+s) \frac{1}{(k^2-s^2)^{2}}
+ \frac{2}{3} R_{\mu,\nu} k^{\mu} k^{\nu} (\gamma^{a}  k_{a}+s) 
\nonumber \\
&&\times \frac{1}{(k^2-s^2)^{3}} - \frac{1}{2} \gamma^{a} J^{c d} R_{c d a \mu} 
k^{\mu} \frac{1}{(k^2-s^2)^{2}}]
\end{eqnarray} 
 where $ J^{a\, b}= \frac{1}{4} [ \gamma^{a} , 
\gamma^{b} ] $ and latin indices are  vierbein indices.

The effective potential is divergent in two and four dimensions and is
finite in three dimensions in the leading 1/N expansion. The four fermion
theory is renormalizable in 2-D flat space.
Using the renormalization condition $ \frac{\partial^{2} V_{0} (\sigma)}{\partial\sigma^{2}}
 \Bigg\vert_{\sigma=M} = \frac{M^{D-2}}{\lambda_{r}}$
where M is the renormalization scale.
The renormalised effective potential in D dimensions is given by \cite{inagaki3}
\begin{eqnarray}
\frac{ V(\sigma,0)}{M^{D}} &=& \frac{1}{2 \lambda_{r}} \frac{\sigma^{2}}{M^{2}}
+\frac{Tr \mathbf {1}}{ 2 \,(4 \pi)^{\frac{D}{2}}} \,\,(D-1)\, \Gamma\,(1-\frac{D}{2}) \,\,
\frac{\sigma^{2}}{M^{2}}
\nonumber \\
&& - \frac{Tr \mathbf{1}}{ (4 \pi)^{\frac{D}{2}} D} \,
\Gamma\,(1-\frac{D}{2}) \,\,\frac{\sigma^{D}}{M^{D}}-
 \frac{Tr \mathbf{1}}{(4 \pi)^{\frac{D}{2}}} \frac{R}{M^2} \frac{1}{ 24} 
\nonumber \\
&&\times \Gamma\,(1-\frac{D}{2}) \,\,\frac{\sigma^{D-2}}{M^{D-2}} 
\end{eqnarray}

We now obtain the four dimensional limit of the NJL model in the MS 
renormalization scheme given by
\begin{eqnarray}
\frac{V (\sigma,0)}{M^{4}} &=& \frac{1}{2 \lambda} \,\,(\frac{\sigma}{ M})^{2}
 -\frac{1}{4 \,\pi^{2}}\,\, ( 1+3\,\, ln \,4 \pi-3 \, \gamma) \,\,
(\frac{\sigma}{M})^{2}
\nonumber \\
&& -\frac{1}{8 \,\pi^{2}}\,\, (\,ln\,\, (\frac{ \sigma}{M})^{2}-\frac{3}{2}-ln \,4 \pi+\gamma) \,\, (\frac{\sigma}{M})^{4}
\nonumber \\
&& -\frac{R}{48 \,M^{2} \pi^{2}} \,\,(\,\,ln \,(\frac{ \sigma}{M})^{2} -1
 -ln\,\, 4 \pi+\gamma)\,\,(\frac{\sigma}{M})^{2}
\end{eqnarray}
Alternatively one could  regularize the divergent part by cutting off the momentum
integral at finite cutoff $\Lambda$ \cite{inagaki2}.
This gives
\begin{eqnarray}
V (\sigma,0) &=& \frac{\sigma^{2}}{2 \lambda} -\frac{1}{(4 \pi)^{2}}\,\,
 [\,\, \sigma^{2} 
\Lambda^{2} + \Lambda^{4}\, ln\, (1+\frac{\sigma^{2}}{\Lambda^{2}}) -\sigma^{4}\,
 ln\, (1+ \frac{\Lambda^{2}}{\sigma^{2}})]
\nonumber \\
&&-\frac{1}{ (4 \pi)^{2}}\, \frac{R}{6} \,\, [\,\, -\sigma^{2} \, 
ln \, (1+ \frac{\Lambda^{2}}{\sigma^{2}})
+ \frac{\Lambda^{2}\sigma^{2}}{\Lambda^{2} +\sigma^{2}}\,\,]
\end{eqnarray}
The two expression can be shown to be equivalent after carrying out the 
renormalization of the coupling constant.
                     The ground state of the theory is determined by the minimum
of the effective potential (10) namely, by solving the gap equation
\begin{equation}
\frac{\partial V(\sigma,0)}{\partial \sigma}\big\vert_{\sigma=m}=0
\end{equation}
For $ \lambda_{r} >\lambda_{c}=\frac{2 \pi^{2}}{1 + 3 ln \,4 \,\pi-3\,\gamma}$,
the minimum of the effective potential is located at the nonvanishing $\sigma$,
the chiral symmetry is broken down dynamically and a dynamical fermion mass is
generated. At the critical point the effective potential has two degenerate
local minima obtained by putting
\begin{equation}
V(\sigma,0)=V(m,0)=0
\end{equation}

                      The solution of the gap equation and the value of the
 critical curvature  can be obtained numerically.
First, we fix the coupling constant $\lambda_{r}$ greater than the critical value
$\lambda_{c}$ corresponding to the  broken symmetric phase.
To study the phase structure in curved space-time we evaluate the effective 
potential  (10) by varying the space-time curvature. We see from Fig.1a
 that the chiral symmetry is restored  as R is increased 
for a fixed $\lambda$. The phase transition induced by curvature is of 
first order as can be seen from Fig.1b. Similar results were obtained by the 
authors of ref. \cite{elizalde1} who showed that a first order phase transition
is induced by curvature in 3-dimensional as well.
Phase structure of NJL model in curved space time with non-trivial topology
has been discussed in the literature. It has been shown that
the combined effect of topology and curvature on phase transition from
chirally symmetric to chirally broken phase is such as to make the transition
first order with the growth in curvature as compared to a second order phase
transition at zero curvature \cite{elizalde3}.
\end{section}
\begin{section}{Phase Structure at Finite Temperature and Chemical Potential
with Curvature}
Four fermion theories in flat space at finite temperature T and chemical potential
 $\mu$ 
in arbitrary dimension has been investigated [5] in the leading order of the 
1/N expansion. The theory under consideration is renormalisable below four 
dimensions and gives an insight into the phase structure of the theory in 
four dimensions.
Following the standard procedure of the Matsubara frequency sums [12] we 
calculate the effective potential in our theory in the leading order of the 
1/N expansion by replacing the phase space integral in eqs (7) by
\begin{equation}
\int \frac{d^4 k_{e}}{(2 \pi)^4} \longrightarrow i \frac{1}{ \beta}
 \sum_{n=0}^\infty \frac{d^3 \bf{k_{e}}}{(2 \pi)^3}   
\end{equation} 
where the four momentum $k^{\mu}$ is given by 
\begin{equation}
k^{\mu} = (i w_{n}-\mu, \bf{k})
\end{equation} 
and the discrete variable $w_{n}$ is given by $ \frac{(2n+1) \pi}{\beta}$

\begin{eqnarray}
V^{\beta}( \sigma, 0) &=& \frac{\sigma^{2}}{2 \lambda} + i \int_{0}^{\sigma} s d s \int
 \frac{d^4 k}{(2 \pi)^4} [ \frac{1}{k^2-s^2}-\frac{R}{12}  \frac{1}{(k^2-s^2)^{2}}
\nonumber \\
&& + \frac{R}{6} \frac{k^{2}}{(k^2-s^2)^{3}}] + \frac{i}{\beta} \sum_{n=0}^\infty
\int_{0}^{\sigma} s d s  \frac{d^3  \bf{{k}}}{(2 \pi)^3}
[ \frac{1}{k^2-s^2}
\nonumber \\
&& - \frac{R}{12}  \frac{1}{(k^2-s^2)^{2}} + \frac{R}{6} 
\frac{k^{2}}{(k^2-s^2)^{3}}]
\end{eqnarray}
\begin{equation}
\nonumber
V^{\beta}( \sigma, 0) = V_{0} (\sigma) +V_{R}(\sigma)+V_{\beta}(\sigma)
\end{equation}

where

\begin{eqnarray}
V_{\beta} (\sigma) &=& \frac{i}{\beta} \sum_{n=0}^\infty
\int_{0}^{\sigma} s d s  \frac{d^3 \bf{k}}{(2 \pi)^3}
[ \frac{1}{k^2-s^2}
 - \frac{R}{12}  \frac{1}{(k^2-s^2)^{2}} 
\nonumber \\
&&+ \frac{R}{6} 
\frac{k^{2}}{(k^2-s^2)^{3}}]
\end{eqnarray} 
Performing integration over s, we get

\begin{eqnarray}
V_{\beta} (\sigma) &=&\frac{i}{2 \beta} \int  \frac{d^3 \bf{k}}{(2 \pi)^3}
\sum_{n=-\infty}^\infty \ln [(w_{n}+i \mu)^{2}+E_{k}^{2}] +  \frac{i}{24}
\frac{R \sigma^{2}}{\beta}  \int \frac{d^3 \bf{k}}{(2 \pi)^3} 
\nonumber \\
&&\sum_{n=-\infty}^\infty
\frac{1}{ [(w_{n}+i \mu)^{2}+E_{k}^{2}]^{2}}
-\frac{i}{12}
\frac{R }{\beta}  \int \frac{d^3 \bf{k}}{(2 \pi)^3} 
\nonumber \\
&& \sum_{n=-\infty}^\infty
\frac{1}{ [(w_{n}+i \mu)^{2}+E_{k}^{2}]}
\end{eqnarray} 
where $E_{k}^{2}=\bf{k}^{2}+s^{2}$ and throwing away the $\beta$ independent
terms and carrying out the summation over n by the standard techniques
we get 
\begin{eqnarray}
V_{\beta} (\sigma) &=& \frac{-2}{ \beta \pi^{2}} \int k^{2} d k
 [ln (1+e^{- \beta [\sqrt {(k^{2}+\sigma^{2})}- \mu] } +\mu \rightarrow -\mu]
\nonumber \\
&& +\frac{R \sigma^{2 }}{12 \pi^{2}} \int \frac{k^{2} d k}
{(k^{2}+\sigma^{2})^{ \frac{3}{2}}} 
(\frac{1}{1+e^{\beta [ \sqrt {(k^{2} + \sigma^{2})} -\mu ]}} +\mu \leftrightarrow -\mu)
\end{eqnarray} 
The effective potential $V^{\beta}(\sigma,0)$ at finite temperature, density and
curvature is then given by 
\begin{eqnarray}
V^{\beta} (\sigma,0) &=& \frac{\sigma^{2}}{2 \,\,\lambda} -\frac{1}{4 \pi^{2}}
\,\, (\, 1+3 \,ln \, 4 \pi- 3 \,\gamma)\,\, \sigma^{2}
\nonumber \\
&& -\frac{1}{8 \pi^{2}} (\,ln \frac{ \sigma^{2}}{M^{2}}-\frac{3}{2}
-ln \,4 \pi+\gamma \,)  \sigma^{4}
\nonumber \\
&& -\frac{R}{48 \pi^{2}} (\, ln \,\frac{ \sigma^{2}}{M^{2}} -1 
-ln \,4 \pi+\gamma) \sigma^{2}
\nonumber \\
&&  \frac{-2}{ \beta \pi^{2}} \int k^{2} d k
 [ln (1+e^{- \beta [\sqrt{ {(k^{2}+\sigma^{2})}}- \mu] } +\mu \rightarrow -\mu]
\nonumber \\
&& +\frac{R \sigma^{2 }}{12 \pi^{2}} \int \frac{k^{2} d k}
{(k^{2}+\sigma^{2})^{ \frac{3}{2}}} ( \frac{1}{ 1+e^{\beta 
[ \sqrt{ (k^{2} + \sigma^{2})} -\mu ]}}+ \mu \leftrightarrow -\mu)
\nonumber \\
&&+\frac{R \sigma^{2 } \beta }{6 \pi^{2}} \int \frac{k^{2} d k}
{(k^{2}+\sigma^{2})} (  \frac{ e^{\beta[ \sqrt {( k^{2}+\sigma^{2 }}-\mu)]}}
{ ( 1+e^{ \beta [ \sqrt{ (k^{2}+\sigma^{2})}- \mu]})^{2} } \mu \leftrightarrow -\mu)
\end{eqnarray} 

At high densities and low temperatures, the matter is degenerate and
the fermion distribution function can be approximated by the step function.
The integral in (20) can be carried out analytically and we obtain

\begin{eqnarray}
\frac{V^{\beta}( \sigma,0)}{M^{4}} &=& \frac{\sigma^{2}}{2 \lambda M^{2}}
 -\frac{1}{4 \pi^{2}} 
\,( 1+3 \,ln \,4 \pi-3 \, \gamma) \,\, \frac{\sigma^{2}}{M^{2}}
\nonumber \\
&& -\frac{1}{8 \pi^{2}} \,\, (ln \, \frac{ \sigma^{2}}{M^{2}}
-\frac{3}{2}-ln \, 4 \pi
+\gamma)  \,\, \frac{\sigma^{4}}{M^{4}}
\nonumber \\
&& -\frac{R}{48 M^{2}\pi^{2}} \, (ln \, \frac{ \sigma^{2}}{M^{2}}
 -1 -ln \, 4 \pi+\gamma) \,\, \frac{\sigma^{2}}{M^{2}}
\nonumber \\
&& -\frac{2}{3 \pi^{2}} \,\,\frac{1}{M^{4}} \,\,[\,\, \frac{\mu}{4} \,  (\mu^{2}-\sigma^{2})^{\frac{3}{2}}
-\frac{3}{8} \sigma^{2} \, \mu \, (\mu^{2}-\sigma^{2})^{\frac{1}{2}}
\nonumber \\
&&+\frac{3}{8}  \, \sigma^{4} \, ln \,\, (\frac{ \mu + 
\sqrt {(\mu^{2}-\sigma^{2})}}{\sigma})]
 +\frac{R \sigma^{2}}{12 M^{4}\pi^{2}} \, [-\frac{ \sqrt{(\mu^{2}-\sigma^{2})}}{\mu}
\nonumber \\
&&+ln \,\, (\frac{ \mu + \sqrt {(\mu^{2}-\sigma^{2})}}{\sigma}) \,\,]
\end{eqnarray}

But at high  temperatures and low densities relevant to the study of
phase transition in the early universe
the integrals in eqs (20) can be done analytically by the usual
tecniques \cite{dolan} and we obtain 

\begin{eqnarray}
\frac{V^{\beta} (\sigma,0)}{M^{4}} &=& \frac{1}{2 \lambda} (\frac{\sigma}{M})^{2}
-\frac{1}{4 \pi^{2}} 
\,( 1+3 \,ln \,4 \pi-3 \, \gamma) \,\, \frac{\sigma^{2}}{M^{2}}
\nonumber \\
&& -\frac{1}{8 \pi^{2}} \,\, (ln \, \frac{ \sigma^{2}}{M^{2}}
-\frac{3}{2}-ln \, 4 \pi
+\gamma)  \,\, \frac{\sigma^{4}}{M^{4}}
\nonumber \\
&& -\frac{R}{48 M^{2}\,\,\pi^{2}} \, (ln \, \frac{ \sigma^{2}}{M^{2}}
 -1 -ln \, 4 \pi+\gamma) \,\, \frac{\sigma^{2}}{M^{2}}
\nonumber \\
&& - \frac{-2}{ \pi^{2} \beta^{4} \,\,M^{4} } [ \frac{7}{360} \pi^{4}- \frac{ \sigma^{2}
\beta^{2} \pi^{2}}{24} + \frac{ \sigma^{4} \beta^{4}}{32} ln \sigma^{2} \beta^{2}
+\frac{\sigma^{4} \beta^{4}}{32} c ]
\nonumber \\
&&+ \frac{R \sigma^{2}}{ 48 \pi^{2}} ( ln \frac{\beta \sigma}{\pi} +\gamma)
\end{eqnarray}
where c=$\frac{3}{2} + 2 \,\,ln 4 \,\,\pi-2 \gamma$. 
We plot the effective potential(20) as a function of $\sigma$
 for different values of R at fixed temperature and density.
We observe from Fig.2 that at finite temperature  symmetry is 
restored at lower values of curvature as compared to the zero temperature case.
The solution of the gap equation corresponds to the dynamical mass of
the fermion and is plotted in Fig.3a at zero chemical potential as a function of curvature
for different values of temperature. We observe that with the increase
in temperature, chiral symmetry is restored at lower values of curvature.
In Fig.3b, we show that solution of the gap equation as a function of 
chemical potential for different values of curvature at T=0. We again find that the symmetry is restored at lower density (lower chemical potential)
as the curvature increases.
\end{section}
\begin{section}{Conclusion}
In this paper we have investigated the phase structure of NJL model in curved
space time using 1/N expansion and working in the linear curvature approximation.
But in one discussion we considered rather large values of curvature and it
would appear that the results obtained may not be reliable. However, it has
been pointed out \cite{inagaki3} that since terms quadratic and/or higher in R are not
divergent, $R^{2}$ terms are expected to be relatively suppressed compared to
linear terms. It has also been demonstrated in de-sitter space 
\cite{elizalde2,muta1} and  Einstein universe \cite{inagaki5} that the results
 obtained in linear approximation are indeed the exact results. A distinguishing
feature of our investigation is that in the presence of external gravity with
positive R, the chiral symmetry restoring phase transition is first order
 even as in flat
space-time, the transition is second order with temperature and becomes first 
order in the presence of density.
In Fig.4 we have shown the phase boundary in temperature, chemical potential 
plane for different values of curvature. We notice that with
 the increase in R, phase transition takes place at lower temperature and
density. In Fig.5 we have shown the phase boundary curve in temperature,
chemical potential plane.  
\end{section}
\begin{section}*{Acknowledgements}
We would like to thank Professor J.V. Narlikar for providing hospitality at
the Inter-University Centre for Astronomy and Astrophysics, Pune 411
007, India where this work was initiated. 
\end{section}

\begin{figure}[ht]
\vskip 15truecm
\includegraphics{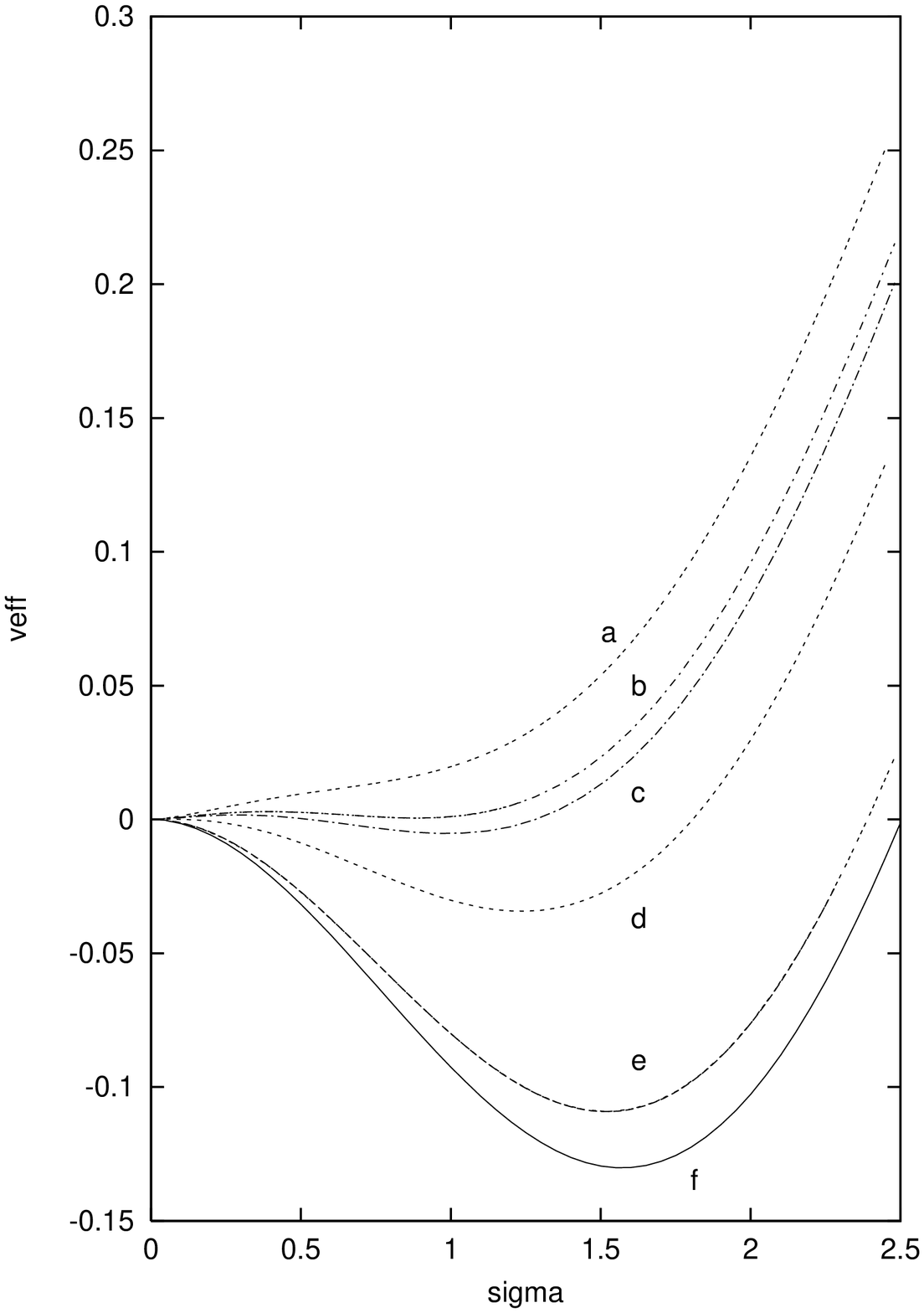}
\caption{Behaviour of Effective Potential $\frac{V}{M^{4}}$ as a function of
$\frac{\sigma}{M}$ for different values of the curvature for $\lambda > 
\lambda_{cr} ( \lambda=10)$. The curves a, b, c, d, e and f are 
for R = 16, 13, 12, 8, 0 and -2 respectively.}
\end{figure}

\begin{figure}[ht]
\vskip 15truecm
\includegraphics{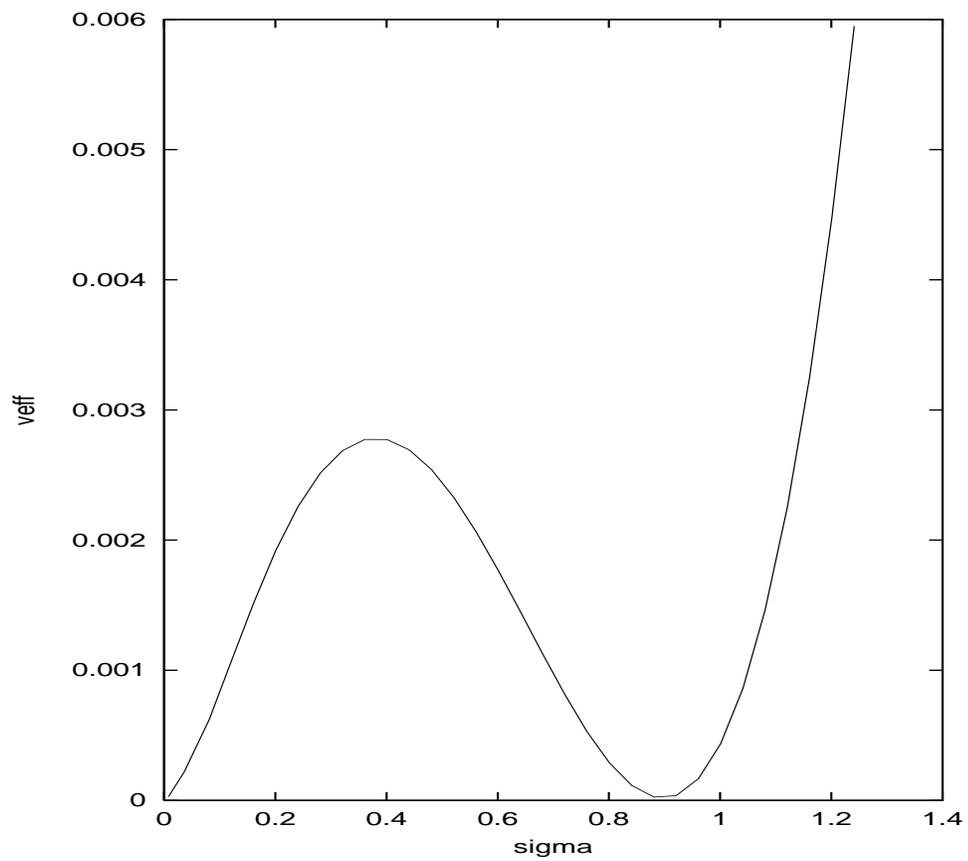}
\caption{Effective potential $\frac{V}{M^{4}}$ for R/$M^{2}$ =12.91 showing the first order phase transition.}
\end{figure}

\begin{figure}[ht]
\vskip 15truecm
\includegraphics{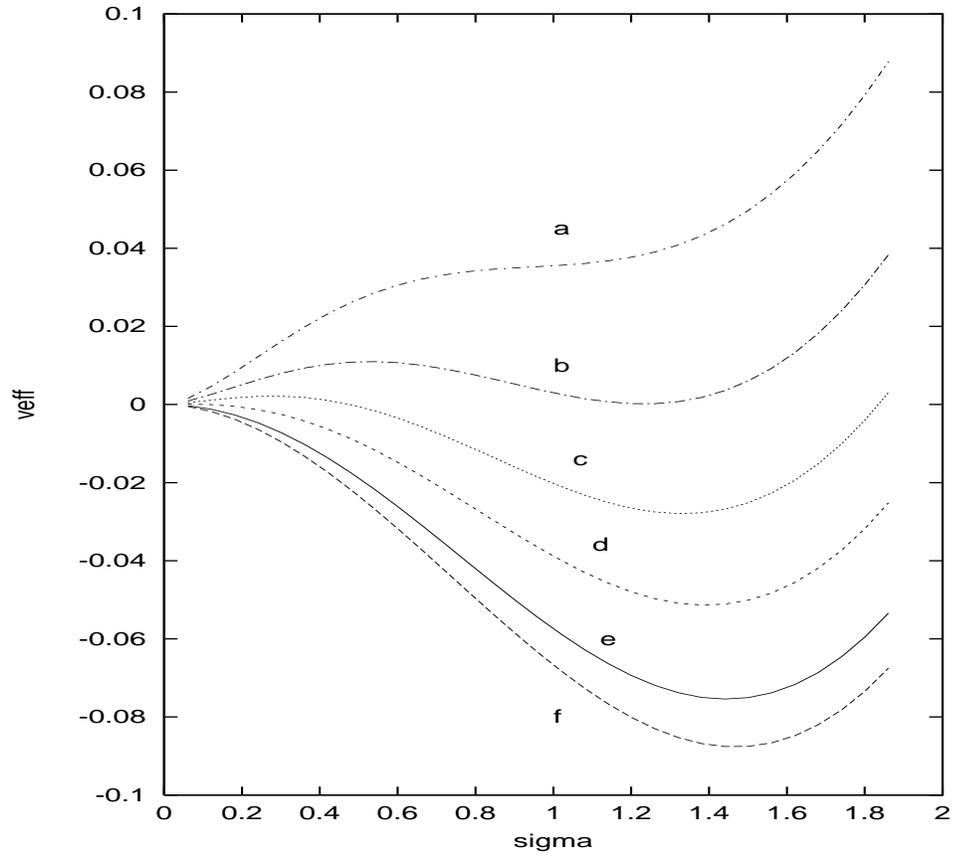}
\caption{Behaviour of effective potential as a function of $\sigma$/M for
different values of the curvature at fixed temperature T/M =0.5. The curves 
a, b, c, d, e and f are for R/M = 10, 6.5, 4, 2, 0, -1 respectively.}
\end{figure}

\begin{figure}[ht]
\vskip 15truecm
\includegraphics{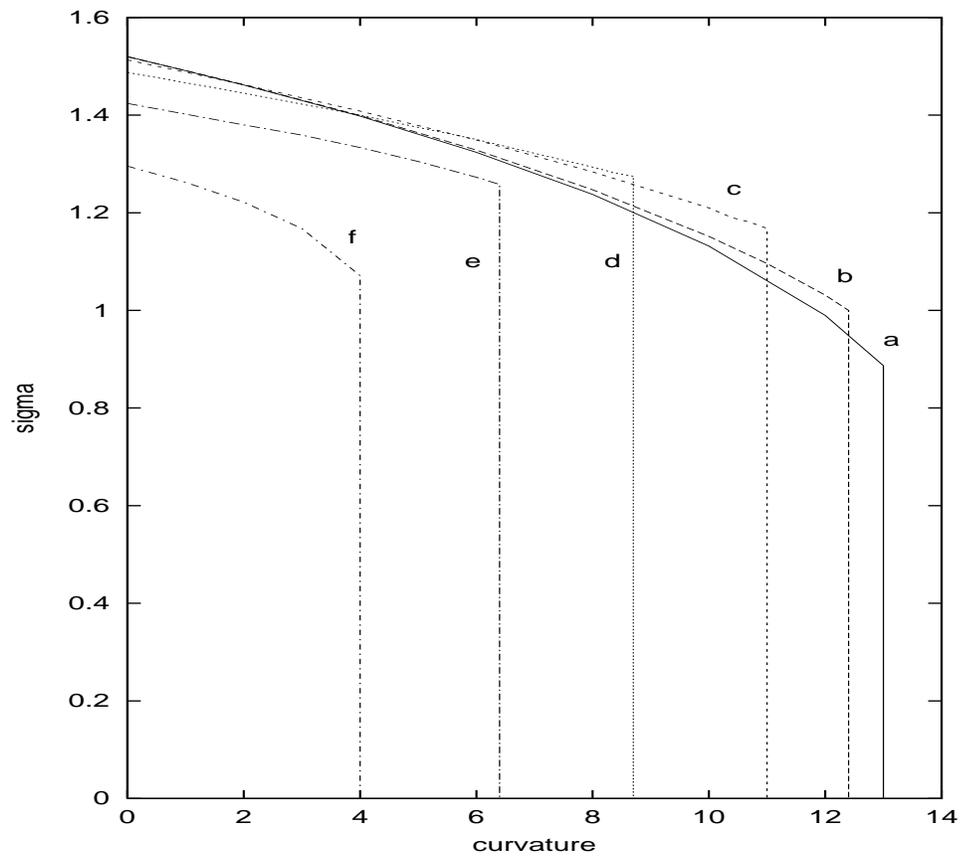}
\caption{Sigma vs curvature for different values of temperature. The
curves a, b, c, d, e and f are for $\frac{T}{M}$ = 0, 0.2, 0.3, 0.4, 0.5, and 0.6
 respectively.}
\end{figure}

\begin{figure}[ht]
\vskip 15truecm
\includegraphics{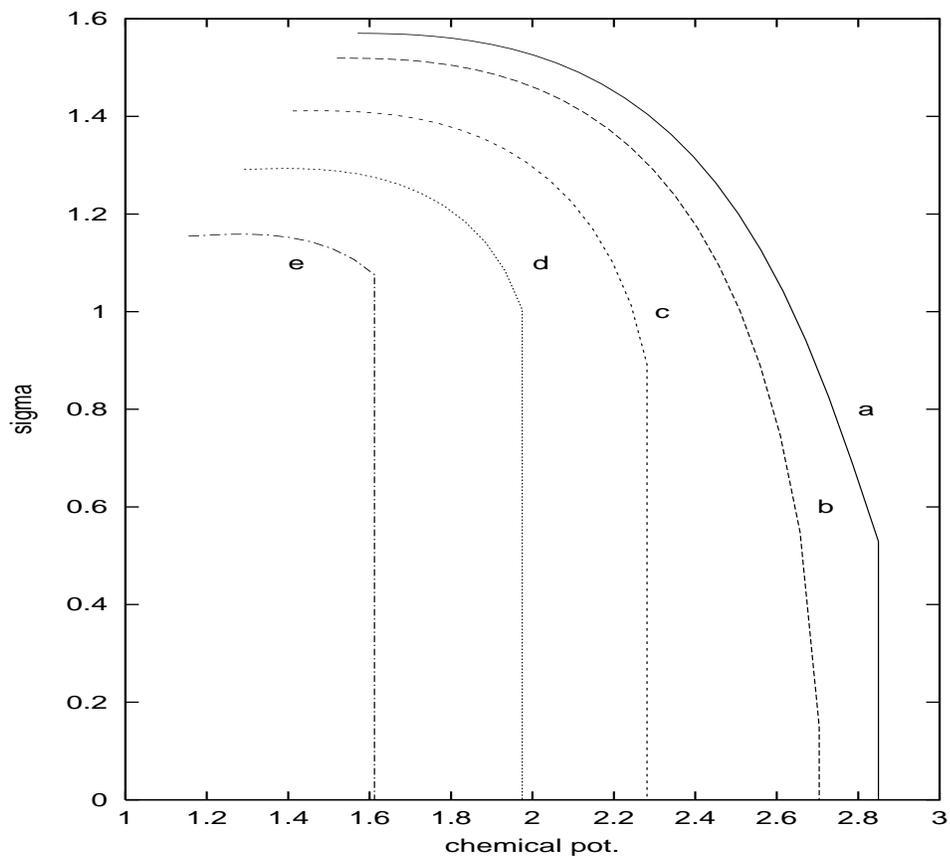}
\caption{Sigma vs chemical potential for different values of curvature. The
curves a, b, c, d and e  are for R/$M^{2}$ = 0, 2, 4, 6, 8 and 10 respectively.}
\end{figure}

\begin{figure}[ht]
\vskip 15truecm
\includegraphics{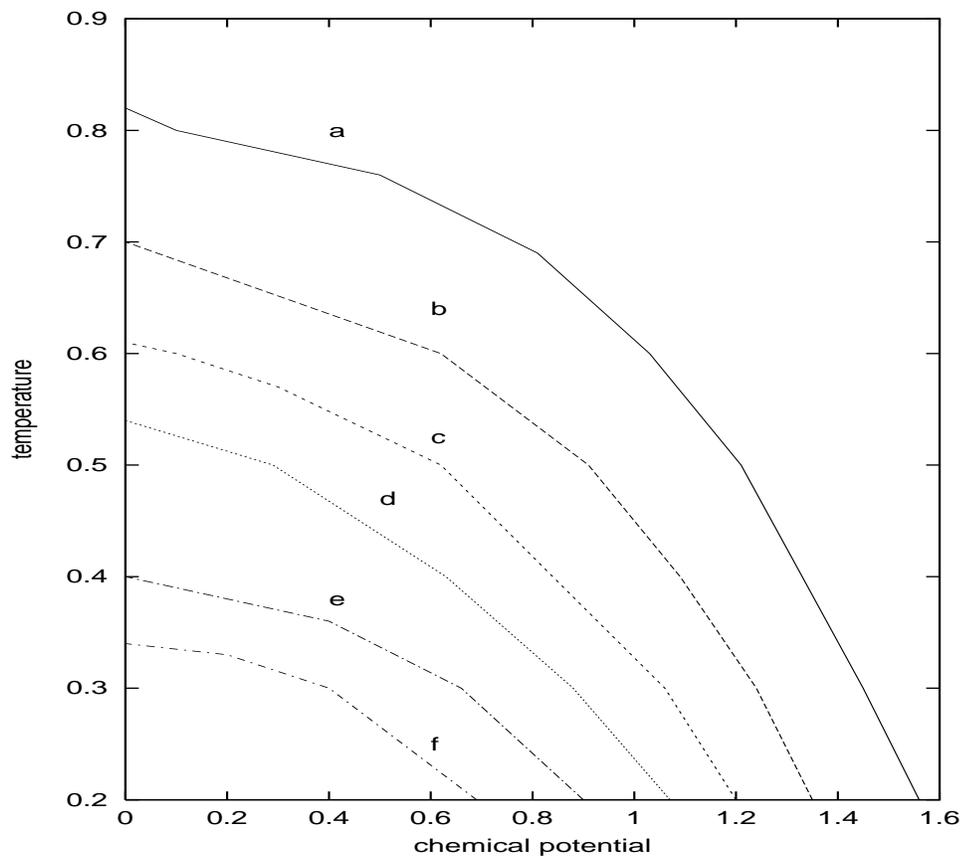}
\caption{Temperature vs chemical pot. for different values of curvature. The
curves a, b, c, d, e and f are for R/$M^{2}$ = 0, 2, 4, 6, 8 and 10 respectively.}
\end{figure}

\begin{figure}[ht]
\vskip 15truecm
\includegraphics{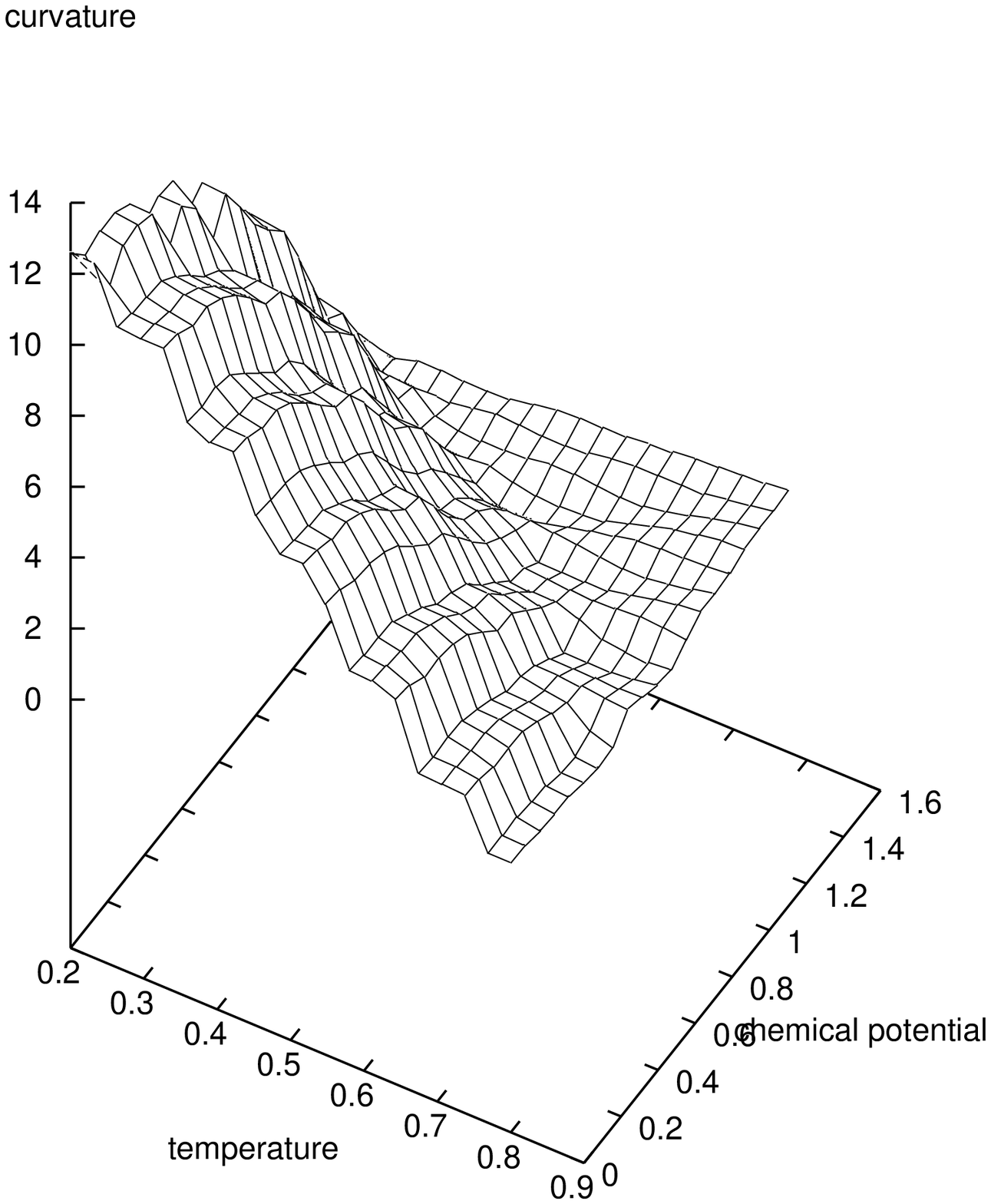}
\caption{Phase diagram in the temperature, curvature and chemical potential plane.}
\end{figure}

\end{document}